\documentclass[11pt,a4paper]{article}
\pdfoutput=1

\usepackage[inline]{enumitem}
\usepackage{etex}
\usepackage{comment}
\usepackage{amsthm}
\usepackage{geometry}
\usepackage{dcolumn}
\usepackage{epsf}
\usepackage{mathrsfs}
\usepackage{multirow}
\usepackage{booktabs}
\usepackage{tabularx}
\usepackage{array}
\usepackage{slashed}
\usepackage{float}
\usepackage{leftidx}
\usepackage{setspace}
\usepackage{verbatim}
\usepackage{adjustbox}
\usepackage{tikz}
\usepackage[caption=false]{subfig}
\usepackage{arydshln}
\usepackage{jheppub}

\numberwithin{equation}{section}
\usetikzlibrary{arrows,decorations.markings,shapes.arrows,patterns,positioning}

\newcommand{\bea}{\begin{eqnarray}}
\newcommand{\eea}{\end{eqnarray}}
\newcommand{\bean}{\begin{eqnarray*}}
\newcommand{\eean}{\end{eqnarray*}}

\def\W #1{\widetilde{#1}}

\def\Label#1{\label{#1}%
  ~~\smash{\hbox to0pt{\raise2ex\hbox{\tiny[#1]}\hss}}}
\def\Label#1{\label{#1}}
\renewcommand{\eqref}[1]{eq.~(\ref{#1})}

\newcommand{\ctobedelete}[1]{}

\newcolumntype{L}[1]{>{\raggedright\let\newline\\\arraybackslash\hspace{0pt}}m{#1}}
\newcolumntype{C}[1]{>{\centering\let\newline\\\arraybackslash\hspace{0pt}}m{#1}}
\newcolumntype{R}[1]{>{\raggedleft\let\newline\\\arraybackslash\hspace{0pt}}m{#1}}

\hypersetup{pageanchor=false}
\title{Direct Evaluation of $n$-point single-trace MHV amplitudes in 4d Einstein-Yang-Mills theory using the CHY Formalism}
\author[a]{Yi-Jian Du,}
\emailAdd{yijian.du@whu.edu.cn}
\author[c]{Fei Teng}
\emailAdd{Fei.Teng@utah.edu}
\author[b,c]{and Yong-Shi Wu}
\emailAdd{wu@physics.utah.edu}
\affiliation[a]{Center for Theoretical Physics,
School of Physics and Technology,
Wuhan University,\\
299 Bayi Road, Wuhan 430072,
China}
\affiliation[b]{Department of Physics and
Center for Field Theory and Particle Physics, Fudan University,\\
220 Handan Road, Shanghai 200433, China}
\affiliation[c]{Department of Physics and Astronomy, University of Utah,\\ 115 South 1400 East, Salt Lake City, UT 84112, USA}

\abstract{In this paper we extend our techniques, developed in a previous paper~\cite{Du:2016blz} for direct evaluation of arbitrary $n$-point tree-level MHV amplitudes in 4d Yang-Mills and gravity theory using the Cachazo-He-Yuan (CHY) formalism, to the 4d Einstein-Yang-Mills (EYM) theory. Any single-trace color-ordered $n$-point tree-level MHV amplitude in EYM theory, obtained by a direct evaluation of the CHY formula, is of an elegant factorized form of a Parke-Taylor factor and a Hodges determinant, much simpler and more compact than the existing formulas in the literature. We prove that our new expression is equivalent to the conjectured Selivanov-Bern-De Freitas-Wong (SBDW) formula, with the help of a new theorem showing that the SBDW generating function has a graph theory interpretation. Together with Ref.~\cite{Du:2016blz}, we provide strong analytic evidence for hidden simplicity in quantum field theory.}

\begin{document}
\maketitle
\hypersetup{pageanchor=true}

\section{Introduction}

The hints about mysterious simplicity of on-shell scattering amplitudes of Yang-Mills fields and gravity have been studied for decades, and new formulations beyond Feyman diagrams were proposed. Among these progresses, the famous Parke-Taylor formula~\cite{Parke:1986gb}, which was proposed in 1986 in terms of the spinor-helicity formalism~\cite{Xu:1986xb}, provides a simple expression for  maximally-helicity-violating (MHV) Yang-Mills amplitudes at tree-level. Gravity amplitudes have more complicated structure than the Yang-Mills ones. Nevertheless, one can use Kawai-Lewellen-Tye (KLT)~\cite{Kawai:1985xq} relation to compute tree-level gravity amplitudes with tree-level Yang-Mills amplitudes as input. In this way, Berends, Giele and Kuijf (BGK) proposed a general formula~\cite{Berends:1988zp} for MHV gravity amplitudes at tree level. Compared with the Parke-Taylor formula for Yang-Mills amplitudes, the BGK formula still presents a complicated expression. On the other hand, tree-level MHV gravity amplitudes are expected to be much simpler. In 2009, Nguyen, Spradlin, Volovich and Wen (NSVW) suggested a diagrammatic construction of MHV gravity amplitudes~\cite{Nguyen:2009jk}. About three years later, the Hodges formula~\cite{Hodges:2012ym}, which expresses tree-level MHV gravity amplitudes by a compact determinant, was established and refreshed our knowledge on the potential simplicity of gravity amplitudes. Actually, the Hodges formula and the NSVW formula are equivalent to each other, as proved by Feng and He~\cite{Feng:2012sy}.

We note that all the above-mentioned formulas for $n$-point MHV amplitudes have been verified using Berends-Giele~\cite{Berends:1987me} and/or Britto-Cachazo-Feng-Witten (BCFW)~\cite{Britto:2004ap, Britto:2005fq} recursive relations (for example, see~\cite{Berends:1987me,Britto:2004ap}). On the other hand, the results produced by recursive relations do not often have a simple and compact form. Therefore, to fully reveal the potential hidden simplicity in scattering amplitudes, one still calls for new formalism that enables direct calculation. Progress in this direction began to appear recently.

In 2013, Cachazo, He and Yuan (CHY) {proposed} a brand-new compact {formula} in a series of work~\cite{Cachazo:2013gna,Cachazo:2013hca,Cachazo:2013iea} for on-shell amplitudes of massless particles, including both gluons and gravitons, {in arbitrary dimensions and helicity configurations}. This initiates a new {perspective} for studying scattering amplitudes in quantum field theories. The CHY formula is based on the scattering equations {for external momenta}, and the polarizations of all external particles are packaged into a reduced pfaffian. {Much effort has} been made to understand this new formula, including the BCFW recursive proof~\cite{Dolan:2013isa}, {study} of the solutions to the scattering  equations~\cite{Monteiro:2013rya,Kalousios:2013eca,Weinzierl:2014vwa,Lam:2015sqb,Cardona:2015ouc,Huang:2015yka,Dolan:2015iln,Sogaard:2015dba,Cardona:2015eba}, the relation to Feynman diagrams and {the method of integrating CHY formula (reducing CHY integrands)}~\cite{Cachazo:2015nwa,Baadsgaard:2015ifa,Lam:2016tlk,Huang:2016zzb,Bjerrum-Bohr:2016juj,Bosma:2016ttj,Zlotnikov:2016wtk,Cardona:2016gon}, the relation to  world-sheet theories \cite{Mason:2013sva,Bjerrum-Bohr:2014qwa}, the extensions to loop amplitudes~\cite{Geyer:2015bja,Adamo:2013tsa,Casali:2014hfa,Adamo:2015hoa,Ohmori:2015sha,Baadsgaard:2015hia,He:2015yua,Geyer:2015jch,Cachazo:2015aol,Feng:2016nrf}, {off-shell case \cite{Lam:2015mgu}}  and other theories~\cite{Cachazo:2014nsa,Cachazo:2014xea,Naculich:2014naa, Naculich:2015zha,Naculich:2015coa,delaCruz:2015raa}.

To us an interesting perspective of the CHY formula would be its analytic computability, resulting in compact and elegant expressions for $n$-point scattering amplitudes (with arbitrary $n$), that should amount to summing up an incredibly huge number of Feynman diagrams.  First, it provides new ways to prove the various, previously conjectured formulas for MHV amplitudes in Yang-Mills and gravity theory, as mentioned in the first paragraph of this section, through direct calculations. Second, this would provide strong analytic evidence for hidden simplicity in quantum field theory. On the other hand, at the present stage of the CHY formalism (or approach), the success of this project would provide a theoretical (quantitative) verification of the validity of the CHY formalism itself, whose simplicity in turn hints on an unfamiliar new formulation of quantum field theory that supersedes usual perturbation theory.

 The authors of the present paper have successfully initiated this project in a previous joint paper~\cite{Du:2016blz} for 4d Yang-Mills theory and Einstein gravity, respectively. The direct evaluation of the CHY formula for $n$-point tree-level MHV amplitudes led to explicitly gauge invariant results,  giving a new proof to the Parke-Taylor and the Hodges formula, respectively, for arbitrary $n$. Two interesting observations are note-worthy. \emph{First}, we found that both the reduced Pfaffian with MHV configuration and the Jacobian determinant in the CHY formula are related to Hodges determinant. \emph{Second}, as conjectured in~\cite{Monteiro:2013rya,Naculich:2014naa} and proved in~\cite{Du:2016blz}, only a special solution \cite{Weinzierl:2014vwa} of scattering equation supports the MHV amplitude. Having these two properties, we expect that the simple formulas for MHV amplitudes, such as the Parke-Taylor formula and the Hodges formula, can be generalized to other theories with a similar CHY integrand. Einstein-Yang-Mills (EYM) theory is such a theory.

EYM is a theory in which the Yang-Mill field minimally couples to gravity. Thus one may expect to find  properties/formulas similar to both Yang-Mills and gravity theory. In~\cite{Selivanov:1997aq, Selivanov:1997ts}, Selivanov proposed a compact formula for single-trace EYM amplitudes in the MHV configuration with two negative helicity gluons. Bern, De Freitas and Wong~\cite{Bern:1999bx} extended KLT relation to EYM theory. As a result, they were able to conjecture a compact formula for the single-trace MHV amplitudes with one negative-helicity gluon and one negative-helicity graviton. This formula is very similar to one of Selivanov such that we will refer both formulas together as Selivanov-Bern-De Freitas-Wong (SBDW) formula in this paper. The  authors of~\cite{Bern:1999bx} further argued that the tree-level single-trace MHV amplitudes with two negative-helicity gravitons should vanish. The SBDW formula associates the amplitudes to the Taylor expansion coefficients of a multivariate generating function. As a result, the calculation involves an ordeal of high-order derivatives while the final expression is neither  explicit nor compact. In~\cite{Cachazo:2014nsa,Cachazo:2014xea}, the CHY formula for EYM amplitudes was proposed. Our motivation is thus to study whether the CHY formalism can give a simpler and more compact expression for EYM amplitudes.

In this paper, we focus on the tree-level single-trace EYM amplitudes in MHV configurations. Based on the CHY formula for EYM theory, we have derived a new compact expression for the MHV amplitudes with at least one negative helicity gluon:
\begin{equation}
M(h_{1}^{+}\cdots i^{-}\cdots g_{{j}}^{-}\cdots g_{r}^{+})\propto\frac{\langle ig_{j}\rangle^{4}}{\langle g_{1}g_{2}\rangle\langle g_{2}g_{3}\rangle\ldots\langle g_{r}g_{1}\rangle}\det(\phi_{\mathsf{h}^{+}})\,,\quad (r\geq 2)\,,
\Label{eq:mainresult}
\end{equation}
where $r$ is the number of gluons (labeled by $g$'s) and the number of gravitons (labeled by $h$'s) is thus $s=n-r$. The above equation is of a factorized form of a Parke-Taylor factor, whose denominator only depends on gluons, and a minor of Hodges matrix, $\det(\phi_{\mathsf{h}^{+}})$, whose indices range within the positive-helicity gravitons only. One of the two negative-helicity particles, say, $g_{j}$, is a gluon, while the other one, say, $i$, can either be a gluon or graviton. We then carry on to prove analytically that the $(h^-h^-)$ MHV amplitudes, in which gluons all carry positive helicities, have to vanish.

In the SBDW formula, the same amplitude as (\ref{eq:mainresult}) is derived from a generating function. We prove a new theorem, showing that such a generating function leads to a weighted sum of spanning forests. It has been shown in~\cite{Feng:2012sy} that the Hodges minor in \eqref{eq:mainresult} has exactly the same graph theory interpretation. Thus we are able to show the equivalence between our new formula and the SBDW prescription for arbitrary $n$.


The structure of this paper is the following. In Sec.~\ref{sec:CHY}, we review the CHY formula in Yang-Mills theory, gravity and EYM theory. In Sec.~\ref{sec:NewEYMFormula}, we derive the new formula for $(g^-g^-)$ and {$(h^-g^-)$} single-trace MHV amplitudes. We also prove that $(h^-h^-)$ amplitudes vanish identically. In Sec.~\ref{sec:ConnectionSBDW}, we provide a graphic representation of the SBDW formula, which is used to prove its equivalence to our new formula. The conclusion of this paper is given in Sec.~\ref{sec:Conclusion}. Finally, we prove that the special solution makes the reduced pfaffian vanish at non-MHV configurations in Appendix~\ref{sec:special}.

\section{Scattering Equations and CHY Formalism}\label{sec:CHY}
The scattering equations for $n$ massless particles are a set of $n$ equations on the complex variables $z_{a}$:
\begin{align}
	&\sum_{\substack{b=1\\b\neq a}}^{n}\frac{s_{ab}}{z_{ab}}=0\,,& a\in\{1,2,\ldots,n\}\,,
\Label{eq:SE}
\end{align}
where $s_{ab}\equiv 2k_{a}\cdot k_{b}$ are the Mandelstam variables and $z_{ab}\equiv z_{a}-z_{b}$. These equations are M\"obius covariant such that we can use this freedom to fix the value of three $z$'s, which implies that among the $n$ equations only $n-3$ of them are linearly independent. The number of solutions to (\ref{eq:SE}) is $(n-3)!$. This fact was first demonstrated by a semi-analytic inductive method in~\cite{Cachazo:2013gna}, and later proved by more elegant algebraic methods in~\cite{Dolan:2015iln,Cardona:2015ouc}. In four dimensions, there are always two special solutions, written in spinor variables:
\begin{align}
	&\sigma_{a}=\frac{\langle a\eta\rangle\langle\theta\xi\rangle}{\langle a\xi\rangle\langle\theta\eta\rangle}\,,& \bar{\sigma}_{a}=\frac{[a\eta][\theta\xi]}{[a\xi][\theta\eta]}\,,\Label{eq:solution}
\end{align}
where the arbitrary projective spinors $\eta$, $\theta$ and $\xi$ encode the M\"obius freedom in the solutions. This spinorial form was first written down by Weinzierl~\cite{Weinzierl:2014vwa}, while it has appeared earlier in other context or forms in~\cite{Roberts:1972,FairlieRoberts:1972,Fairlie:2008dg,Monteiro:2013rya}. For these two solutions, we have:
\begin{align}
	&\sigma_{ab}\equiv\sigma_{a}-\sigma_{b}=\frac{\langle ab\rangle\langle\theta\xi\rangle\langle\eta\xi\rangle}{\langle a\xi\rangle\langle b\xi\rangle\langle\theta\eta\rangle}\,,& &\bar{\sigma}_{ab}\equiv\bar{\sigma}_{a}-\bar{\sigma}_{b}=\frac{[ ab][\theta\xi][\eta\xi]}{ [a\xi] [b\xi][\theta\eta]}\,,
\Label{eq:sigmaab}
\end{align}
which will be used frequently when studying MHV amplitudes. In the following, We will use $\omega_{a}$ for generic solutions to \eqref{eq:SE}, while $\sigma_{a}$ and $\bar{\sigma}_{a}$ are used only for the special solutions (\ref{eq:solution}).

CHY formalism states that generic tree-level $n$-point massless amplitudes are supported only by the solutions to the scattering equations. Namely, it can be calculated by:
\begin{equation}
	A_{n}=\sum_{\{\omega\}\in\,\text{sol.}}\frac{\mathcal{I}_{n}}{{\det}'(\Phi)}\,,
\end{equation}
where $\mathcal{I}_{n}$ is the CHY integrand, defining various theories, and the matrix $\Phi$ is:
\begin{equation}
\renewcommand{\arraystretch}{1.5}
\Phi_{ab}=\left\{\begin{array}{>{\displaystyle}l @{\hspace{1.5em}} >{\displaystyle}l}
\frac{s_{ab}}{\omega_{ab}^{2}} & a\neq b \\
-\sum_{c\neq a}\frac{s_{ac}}{\omega_{ac}^{2}} & a=b \\
\end{array}\right.\,.
\Label{eq:Phi}
\end{equation}
This matrix has rank deficiency three so that to have a nonzero determinant we need to delete three rows $(i,j,k)$ and three columns $(p,q,r)$. With the resultant submatrix $\Phi_{pqr}^{ijk}$, the ${\det}'$ is defined as:
\begin{equation}
   {\det}'(\Phi)\equiv\text{perm}(ijk)\,\text{perm}(pqr)\frac{\det\left(\Phi_{pqr}^{ijk}\right)}{\omega_{ij}\omega_{jk}\omega_{ki}\omega_{pq}\omega_{qr}\omega_{rp}}\,,
\end{equation}
where $\text{perm}(ijk)$ is the signature of the permutation that moves $(1,2,\ldots,n)$ into $(i,j,k,\ldots,n)$, with the $(\ldots)$ keeping the original ascending order. It has been shown that ${\det}'(\Phi)$ is independent of the choice of $(i,j,k)$ and $(p,q,r)$~\cite{Cachazo:2013hca}.

In this work, our main subject is single-trace tree-level EYM amplitudes, whose CHY integrand has a close relation with those of pure Yang-Mills and pure gravity. Next, we are going to give a brief review on the CHY formalism of these theories.

\subsection{Yang-Mills and pure gravity}
The integrands for color-ordered Yang-Mills and pure gravity amplitudes are:
\begin{align}
	&\mathcal{I}_{n}(\{k,\epsilon,\omega\})=\frac{\text{Pf}\,'[\Psi(k,\epsilon,\omega)]}{\omega_{12}\omega_{23}\ldots \omega_{n1}}& &\text{color-ordered Yang-Mills,}\nonumber\\*
	&\mathcal{I}_{n}\left(\{k,\epsilon,\W{\epsilon},\omega\}\right)={\text{Pf}\,'\left[\Psi(k,\epsilon,\omega)\right]\times\text{Pf}\,'\left[\Psi(k,\W{\epsilon},\omega)\right]}& &\text{pure gravity,}
\Label{eq:YMgravity}
\end{align}
where $\{k\}$ is the set of external momenta and $\{\epsilon\}$ (both $\{\epsilon\}$ and $\{\W{\epsilon}\}$) is the set of polarizations for gluons (gravitons). The $2n\times 2n$ antisymmetric matrix $\Psi$ is given by:
\begin{equation}
	\Psi(\{k,\epsilon,\omega\})=\left(\begin{array}{cc}
		A & -C^{T} \\
		C & B \\
		\end{array}\right)\,,
\Label{eq:Psi}
\end{equation}
where the blocks are:
\begin{align}
& A_{ab}=\left\{\begin{array}{>{\displaystyle}c @{\hspace{1em}} >{\displaystyle}l}
\frac{s_{ab}}{\omega_{ab}} & a\neq b\\
0 & a=b \\
\end{array}\right.\,,&
& B_{ab}=\left\{\begin{array}{>{\displaystyle}c @{\hspace{1em}} >{\displaystyle}l}
\frac{2\epsilon_{a}\cdot\epsilon_{b}}{\omega_{ab}} & a\neq b\\
0 & a=b \\
\end{array}\right.\,,&
& C_{ab}=\left\{\begin{array}{>{\displaystyle}l @{\hspace{1em}} >{\displaystyle}l}
\frac{2\epsilon_{a}\cdot k_{b}}{\omega_{ab}} & a\neq b\\
-\sum_{c\neq a}\frac{2\epsilon_{a}\cdot k_{c}}{\omega_{ac}} & a=b \\
\end{array}\right.\,.
\Label{eq:ABC}
\end{align}
The upper half of $\Psi$, $(A,-C^{T})$, has two null vectors such that we need to delete two rows and columns in the first $n$ rows and columns to obtain a nonzero pfaffian. Thus $\text{Pf}\,'$ is defined as:
\begin{equation}
\text{Pf}\,'(\Psi)=\frac{\text{perm}(ij)}{\omega_{ij}}\text{Pf}\,(\Psi_{ {i} {j}}^{ {i} {j}})\,,\Label{eq:rf}
\end{equation}
where $1\leqslant i<j\leqslant n$. It is also independent of the choice of $(i,j)$. Since both $\text{Pf}\,'(\Psi)$ and ${\det}'(\Phi)$ are invariant under permutations, the total Yang-Mills amplitudes can be obtained by
\begin{equation}
	\mathcal{A}_{n}=\sum_{\{\omega\}\in\,\text{sol.}}\left[\frac{\text{Tr}\,(T^{a_{1}}T^{a_{2}}\ldots T^{a_{n}})}{\omega_{12}\omega_{23}\ldots\omega_{n1}}+\text{non-cyclic perm.}\right]\frac{\text{Pf}\,'(\Psi)}{{\det}'(\Phi)}\,,
\end{equation}
where $T^{a_{i}}$ is the Lie algebra generator of the gauge group.

\subsection{Single-trace Einstein-Yang-Mills}
\label{sec:singletrace}
The EYM amplitude, $\mathcal{M}_{s,r}$, is characterized by the number of external gravitons $s$ and gluons $r$, with $s+r=n$. For convenience, we define the set of gravitons and gluons to be $\mathsf{h}$ and $\mathsf{g}$, while the set of external particles be $\mathsf{p}=\mathsf{h}\cup\mathsf{g}=\{1,2,\ldots,n\}$. By convention, we use:
\begin{align*}
	&\mathsf{h}=\{1,2,\ldots,s\}\equiv\{h_1,h_2,\ldots,h_s\}\,,& &\mathsf{g}=\{s+1,s+2,\ldots,s+r\}\equiv\{g_1,g_2,\ldots,g_r\}\,.
\end{align*}
We will also use the sets of $+$ and $-$ helicity gravitons $\mathsf{h}_{\pm}$ and gluons $\mathsf{g}_{\pm}$, as well as $\mathsf{p}_{\pm}=\mathsf{h}_{\pm}\cup\mathsf{g}_{\pm}$. The orders of these sets are denoted as
\begin{align}
	& n=|\mathsf{p}|\,,& & s=|\mathsf{h}|\,,& & r=|\mathsf{g}|\,,\nonumber\\
	& n^{\pm}=|\mathsf{p}_{\pm}|\,,& & s^{\pm}=|\mathsf{h}_{\pm}|\,,& & r^{\pm}=|\mathsf{g}_{\pm}|\,.
\end{align}
In~\cite{Cachazo:2014nsa}, the authors proposed an integrand for single-trace EYM amplitudes, which reads:
\begin{align}
	\mathcal{M}(h_{1},\ldots,h_{s},g_{{1}},\ldots,g_{{r}})&=\sum_{\{\omega\}\in\,\text{sol.}}\frac{\mathcal{I}_{s,r}}{{\det}'(\Phi)}\nonumber\\
	&=\sum_{\{\omega\}\in\,\text{sol.}}\left[\frac{\text{Tr}\,(T^{a_{{s+1}}}\ldots T^{a_{{n}}})}{\omega_{g_{1}g_{2}}\omega_{g_{2}g_{3}}\ldots\omega_{g_{r}g_{1}}}+\text{non-cyclic perm.}\right]\frac{\text{Pf}\,(\Psi_{\mathsf{h}})\text{Pf}\,'(\Psi)}{{\det}'(\Phi)}\,,
\Label{eq:EYM}
\end{align}
where $\Phi$ and $\Psi$ are given by \eqref{eq:Phi} and (\ref{eq:Psi}) respectively, the same as the Yang-Mills case. The $2s\times 2s$ matrix $\Psi_{\mathsf{h}}$ is given by:
\begin{equation}
	\Psi_{\mathsf{h}}(\{k,\W{\epsilon},\omega\})=\left(\begin{array}{cc}
		A_{\mathsf{h}} & -C_{\mathsf{h}}^{T} \\
		C_{\mathsf{h}} & B_{\mathsf{h}} \\
		\end{array}\right)\,,
\Label{eq:Psih}
\end{equation}
where $A_{\mathsf{h}}$, $B_{\mathsf{h}}$ and $C_{\mathsf{h}}$ are $s\times s$ dimensional diagonal submatrices of $A$, $B$, and $C$, whose indices range within the graviton set $\mathsf{h}$. In this formula, we assume that the gluons have the polarization $\epsilon_{a}^{\mu}$ and gravitons have $\epsilon_{a}^{\mu\nu}=\epsilon^{\mu}_{a}\W{\epsilon}^{\nu}_{a}$. We note that if there is only one gluon, namely, $s=n-1$, the amplitudes are identically zero due to $\text{Pf}\,(\Psi_{\mathsf{h}})=0$, independent of helicity configurations and solutions. In this case, $\Psi_{\mathsf{h}}$ can be obtained from $\Psi$ by deleting the $n$-th and $2n$-th row and column. However, as discussed below \eqref{eq:ABC}, there are two null vectors in the upper half of $\Psi$ such that after deleting the $n$-th row and column, there is still one, which makes the pfaffian vanish. Physically, the vanishing of this amplitude is easy to understand from the conservation of color quantum numbers.

Also in~\cite{Cachazo:2014nsa}, the authors have shown that \eqref{eq:EYM} can give the correct soft limit, and checked numerically that the amplitudes agree with the known results. In the next section, we are going to derive analytically the single-trace MHV amplitudes. We will show that they agree with those given by Selivanov, Bern, De Freitas and Wong (SBFW)~\cite{Selivanov:1997aq,Selivanov:1997ts,Bern:1999bx} in Sec.~\ref{sec:ConnectionSBDW}.
\section{Single-Trace MHV Amplitudes for Einstein-Yang-Mills} \label{sec:NewEYMFormula}
At MHV, the EYM amplitudes can be put into three categories:
\begin{enumerate}
\item two negative helicity gluons, hereafter $(g^{-}g^{-})$.
\item one negative helicity graviton and the other gluon, hereafter $(h^{-}g^{-})$.
\item two negative helicity gravitons, hereafter $(h^{-}h^{-})$.
\end{enumerate}
For the first two cases, SBFW~\cite{Selivanov:1997aq,Selivanov:1997ts,Bern:1999bx} have provided a compact formula to calculate the amplitudes, while in Sec.~\ref{sec:gg} and \ref{sec:hg}, we provide a more explicit expression in terms of the Hodges determinant~\cite{Hodges:2012ym}, using the CHY integrand (\ref{eq:EYM}). For the last case, \cite{Bern:1999bx} argued that the amplitude vanishes by imposing the required factorization properties. In Sec.~\ref{sec:hh}, we prove analytically that this is the case.

Since the factor $\text{Pf}\,'(\Psi)$ is shared by both the single-trace EYM integrand (\ref{eq:EYM}) and the Yang-Mills one (\ref{eq:YMgravity}), EYM amplitudes must be supported by the same set of solutions at most, if not less. It has been conjectured that only the special solution $\sigma$ ($\bar{\sigma}$) shown in \eqref{eq:solution} supports the Yang-Mills MHV (anti-MHV, hereafter $\overline{\text{MHV}}$) amplitudes~\cite{Cachazo:2013iaa,Monteiro:2013rya,Naculich:2014naa}. In~\cite{Du:2016blz}, the present authors proved this point analytically, namely, using the CHY prescription, the solutions in \eqref{eq:solution} do reproduce the correct Parke-Taylor formula~\cite{Parke:1986gb} for Yang-Mills and Hodges formula~\cite{Hodges:2012ym} for pure gravity at MHV. In particular, if the two negative helicities are located at position $i$ and $j$, we have
\begin{equation}
  \frac{\text{Pf}\,'[\Psi(\sigma)]}{\det'[\Phi(\sigma)]}=\frac{(-1)^{s(n)}(\sqrt{2}\,)^{n}\langle ij\rangle^{4}}{F^{n}P^{2}_{\xi}}\,,
\Label{eq:Pfdet}
\end{equation}
where $\Psi$ and $\Phi$ are evaluated on $\sigma$ (with $k$ and $\epsilon$ dependency suppressed). 
Next, $F$ and $P_{\xi}$ contain only the M\"obius degrees of freedom:
\begin{align}
	&F=\frac{\langle\theta\eta\rangle}{\langle\eta\xi\rangle\langle\theta\xi\rangle}\,,& &P_{\xi}=\prod_{a=1}^{n}\langle a\xi\rangle\,,
\end{align}
which will not appear in the final expression for amplitudes. Finally, $s(n)=(n^{2}-3n+8)/2$ provides an unobservable overall sign to the amplitudes. On the other hand, the vanishing of $\text{Pf}\,'(\Psi)$ has been proved for the solution $\{\overline{\sigma}_{a}\}$ in (\ref{eq:solution}) while it remains a numerical fact for the others. More generally, there is an intriguing Eulerian number pattern on how $\text{Pf}\,'(\Psi)$ is supported by the solutions. Such an observation has been indicated in some earlier work, for example, \cite{Spradlin:2009qr,Cachazo:2016sdc}, but a full understanding is still elusive.

As an immediate application of the discussion above, since the EYM amplitude is proportional to $\text{Pf}\,'(\Psi)$, at MHV we only need to calculate $\text{Pf}\,(\Psi_{\mathsf{h}})$ at the special solution $\sigma$, although in general, $\text{Pf}\,(\Psi_{\mathsf{h}})$ is nonzero at other solutions.
\subsection{\texorpdfstring{$(g^{-}g^{-})$}{(g-g-)} amplitudes}
\label{sec:gg}
In this case, suppose the two negative helicity gluons are $g_{i}$ and $g_{j}$ while all the gravitons have positive helicities. We choose the polarization $\W{\epsilon}$ to be:
\begin{align}
&\W{\epsilon}^{\mu}_{a}(+)=\frac{\langle q|\gamma^{\mu}|a]}{\sqrt{2}\langle qa\rangle}\,,& &(a\in\mathsf{h})\,,
\end{align}
namely, all the reference vectors are the same. Then we always have $\W{\epsilon}\cdot\W{\epsilon}=0$ such that $B_{\mathsf{h}}$ is identically zero. The $2s\times 2s$ matrix $\Psi_{\mathsf{h}}$ now has the form:
\begin{equation}
	\Psi_{\mathsf{h}}=\left(\begin{array}{cc}
		A_{\mathsf{h}} & -C_{\mathsf{h}}^{T} \\
		C_{\mathsf{h}} & 0
	\end{array}\right)\,.
\end{equation}
Then $\text{Pf}\,(\Psi_{\mathsf{h}})\neq 0$ if and only if $C_{\mathsf{h}}$ is of full rank. Indeed, if $C_{\mathsf{h}}$ has rank deficiency, we can always make one row of it zero by elementary transformations, and thus $\Psi_{\mathsf{h}}$ has one row of zeros such that $\det(\Psi_{\mathsf{h}})=0$. Next, independent of the solutions, we always have:
\begin{equation}
	\text{Pf}\,(\Psi_{\mathsf{h}})=(-1)^{s(s+1)/2}\det({C}_{\mathsf{h}})\,,
\Label{eq:PfPsih}
\end{equation}
namely, we can pretend that $A_{\mathsf{h}}=0$. The reason is that if $C_{\mathsf{h}}$ is of full rank, there always exists an elementary transformation that makes $A_{\mathsf{h}}$ vanish, and then we can use the formula for the pfaffian of an off-diagonal block matrix. After plugging in $\sigma$, we get:
\begin{align}
    C_{ab}&=-\sqrt{2}F\frac{[h_ah_b]\langle h_a\xi\rangle\langle h_b\xi\rangle\langle h_bq\rangle}{\langle h_ah_b\rangle\langle h_aq\rangle}\,,& &C_{aa}=\sqrt{2}F\langle h_a{\xi}\rangle^{2}\sum_{\substack{l=1\\l\neq h_a}}^{{n}}\frac{[h_al]\langle l\xi\rangle\langle lq\rangle}{\langle h_al\rangle\langle h_a\xi\rangle\langle h_aq\rangle}\,.
\end{align}
When calculating the determinant, we can pull out $F\langle h_a\xi\rangle/\langle h_aq\rangle$ from each row and $\langle h_b\xi\rangle\langle h_bq\rangle$ from each column. This calculation leads to
\begin{equation}
	\frac{\text{Pf}\,[\Psi_{\mathsf{h}}(\sigma)]}{\sigma_{g_{1}g_{2}}\sigma_{g_{2}g_{3}}\ldots\sigma_{g_{r}g_{1}}}=(-1)^{s(s-1)/2}(\sqrt{2}\,)^{s}F^{{n}}\left(P_{\xi}\right)^{2}\frac{\det(\phi_{\mathsf{h}})}{\langle g_{1}g_{2}\rangle\langle g_{2}g_{3}\rangle\ldots\langle g_{r}g_{1}\rangle}\,,
\Label{eq:Pfcf}
\end{equation}
where $\phi_{\mathsf{h}}$ is the $s\times s$ diagonal submatrix of the Hodges matrix $\phi$~\cite{Hodges:2012ym} with all gluon rows and columns deleted:
\begin{equation}
  \phi_{\mathsf{h}}\equiv \phi_{g_{1}\cdots g_{r}}^{g_{1}\cdots g_{r}}\,.
\end{equation}
The $n\times n$ Hodges matrix $\phi$ is given by:
\begin{align}
	&\phi_{ab}=\frac{[ab]}{\langle ab\rangle}& &(a\neq b,\quad a,b\in\mathsf{p})\,,& &\phi_{aa}=-\sum_{\substack{l=1\\l\neq a}}^{n}\frac{[al]\langle l\xi\rangle\langle l\eta\rangle}{\langle al\rangle\langle a\xi\rangle\langle a\eta\rangle}& &(a=b,\quad a\in\mathsf{p})\,, \Label{Eq:HodgesMatrix}
\end{align}
where the spinors $|\xi\rangle$ and $|\eta\rangle$ represent gauge freedom and the diagonal elements $\phi_{aa}$ do not depend on it. We can even choose them to be an external momentum spinor $|i\rangle$, as long as the $i$-th row and column have been deleted for some reason. The matrix $\phi$ has rank $n-3$ so that its nonzero minors are at most $n-3$ dimensional.

Combining \eqref{eq:Pfcf} with \eqref{eq:Pfdet}, we get the color-ordered $(g^{-}g^{-})$ MHV amplitude as:\footnote{The ``$\propto$'' sign means that we neglect an overall coefficient which depends only on $n$ and $s$.}
\begin{align}
	M(h_{1}^{+}\cdots h_{s}^{+};g_{{1}}^{+}\cdots g_{{i}}^{-}\cdots g_{{j}}^{-}\cdots g_{{r}}^{+})\propto\frac{\langle g_{i}g_{j}\rangle^{4}}{\langle g_{1}g_{2}\rangle\langle g_{2}g_{3}\rangle\ldots\langle g_{r}g_{1}\rangle}\det(\phi_{\mathsf{h}})\,.
\Label{eq:Mgg}
\end{align}
For $\mathsf{h}=\varnothing$, we define $\phi_{\varnothing}=1$. In Sec.~\ref{sec:ConnectionSBDW}, we will show that $\det(\phi_{\mathsf{h}})$ can arise from the SBFW prescription.
\subsection{\texorpdfstring{$(h^{-}g^{-})$}{(h-g-)} amplitudes}
\label{sec:hg}
In this case, the particle $h_{i}$ and $g_{j}$ have negative helicities while all the other particles have positive helicities. The set of positive helicity gravitions is thus $\mathsf{h}_{+}=\mathsf{h}\backslash\{h_i\}$. We fix the gauge freedom in $\W{\epsilon}$ as:
\begin{align}
	&\W{\epsilon}^{\mu}_{i}(-)=\frac{\langle h_i|\gamma^{\mu}|q]}{\sqrt{2}[qh_i]}\,,& &\W{\epsilon}^{\mu}_{a}(+)=\frac{\langle h_i|\gamma^{\mu}|a]}{\sqrt{2}\langle h_ia\rangle}\,,& &(a\in\mathsf{h}_{+})\,,
\end{align}
such that we still have $B_{\mathsf{h}}=0$ and \eqref{eq:PfPsih} holds. In addition, the $i$-th column of $C_{\mathsf{h}}$ is zero except for the diagonal element $C_{ii}$ under our gauge choice. Therefore, $\det(C_{\mathsf{h}})$ evaluates to
\begin{equation}
\det(C_{\mathsf{h}})=C_{ii}\det\left[(C_{\mathsf{h}})_{i}^{i}\right]=(-\sqrt{2}\,)^{s}F^{s}\left(\prod_{a=1}^{s}\langle h_a\xi\rangle^{2}\right)\det\left[(\phi_{\mathsf{h}})_{i}^{i}\right]\,,
\end{equation}
where in the last equality, we have plugged in the special solution $\sigma$, such that:
\begin{equation}
	C_{ii}=-\sum_{\substack{l=1\\l\neq i}}^{n}\frac{2\epsilon_{i}(-)\cdot k_{l}}{\sigma_{il}}=-\sqrt{2}F\langle h_i\xi\rangle\sum_{\substack{l=1\\l\neq h_i}}^{n}\frac{[lq]\langle l\xi\rangle}{[qh_i]}=-\sqrt{2}F\langle h_i\xi\rangle^{2}\,.
\end{equation}
This calculation leads to
\begin{equation}
\frac{\text{Pf}\,[\Psi_{\mathsf{h}}(\sigma)]}{\sigma_{g_{1}g_{2}}\sigma_{g_{2}g_{3}}\ldots\sigma_{g_{r}g_{1}}}=(-1)^{s(s-1)/2}(\sqrt{2}\,)^{s}F^{{n}}\left(P_{\xi}\right)^{2}\frac{\det[(\phi_{\mathsf{h}})_{i}^{i}]}{\langle g_{1}g_{2}\rangle\langle g_{2}g_{3}\rangle\ldots\langle g_{r}g_{1}\rangle}\,,
\end{equation}
such that the color-ordered amplitude reads:
\begin{align}
	M(h_{1}^{+}\cdots h_{i}^{-}\cdots h_{s}^{+};g_{{1}}^{+}\cdots g_{{j}}^{-}\cdots g_{{r}}^{+})\propto\frac{\langle h_i g_{j}\rangle^{4}}{\langle g_{1}g_{2}\rangle\langle g_{2}g_{3}\rangle\ldots\langle g_{r}g_{1}\rangle}\det[(\phi_{\mathsf{h}})_{i}^{i}]\,.
\Label{eq:Mhg}
\end{align}
Comparing \eqref{eq:Mgg} with (\ref{eq:Mhg}), we find that for at least one negative helicity gluon $g_{j}$:
\begin{align}
	M(h_{1}^{+}\cdots i^{-}\cdots g_{{j}}^{-}\cdots g_{r}^{+})\propto\frac{\langle i g_{j}\rangle^{4}}{\langle g_{1}g_{2}\rangle\langle g_{2}g_{3}\rangle\ldots\langle g_{r}g_{1}\rangle}\det(\phi_{\mathsf{h}_{+}})\,,
\Label{eq:Mgghg}
\end{align}
where $\phi_{\mathsf{h}_{+}}$ is a submatrix of the Hodges matrix $\phi$ whose indices belong to the set of positive helicity gravitons. For $\mathsf{h}_{+}=\varnothing$, we define $\phi_{\varnothing}=1$. {After expanding the determinants, we can check the boundary cases with \emph{only two gluons}:
\begin{itemize}
\item For $(g^-g^-)$ configuration, the amplitude $M(h_1^+,h_2^+,\cdots h_s^+;g_1^-,g_2^-)$ vanishes since now $\phi_{\mathsf{h}_{+}}$ is $n-2$ dimensional while $\phi$ only has rank $n-3$, which agrees with the analysis in~\cite{Chen:2010ct}.
\item For $(h^-g^-)$ configuration, the amplitude $M(h_1^-,h_2^+,\cdots h_s^+;g_1^-,g_2^+)$, matches perfectly with those existing results~\cite{Chen:2010sr,Chen:2010ct} for three- and four-point cases.
\end{itemize}
  }
\subsection{\texorpdfstring{$(h^{-}h^{-})$}{(h-h-)} amplitudes}
\label{sec:hh}
In this case, the two negative helicity particles are $h_{i}$ and $h_{j}$ such that $\mathsf{h}_{-}=\{h_{i},h_{j}\}$. We are going to prove that $\text{Pf}\,[\Psi_{\mathsf{h}}(\sigma)]=0$ such that the amplitude vanishes identically. Actually, here we prove a stronger statement than this: \emph{if $|\mathsf{h}_{-}|\geqslant 2$, we always have} $\text{Pf}\,[\Psi_{\mathsf{h}}(\sigma)]=0$. First, we choose the polarizations as
\begin{align}
	&\W{\epsilon}^{\mu}_{a}(-)=\frac{\langle a|\gamma^{\mu}|q]}{\sqrt{2}[qa]}\,,& &(a\in\mathsf{h}_{-}) &\W{\epsilon}^{\mu}_{a}(+)=\frac{\langle p|\gamma^{\mu}|a]}{\sqrt{2}\langle pa\rangle}\,,& &(a\in\mathsf{h}_{+})\,.
\Label{eq:gaugechoice}
\end{align}
where $p$ and $q$ are two arbitrary reference vectors that do not coincide with any graviton momentum. By plugging in $\sigma$ and extracting common factors in each row and column, we can reach at
\begin{equation}
	\text{Pf}\,[\Psi_{\mathsf{h}}(\sigma)]=F^{s}(-\sqrt{2}\,)^{s}\Bigg(\prod_{a\in\mathsf{h}}\langle a\xi\rangle^{2}\Bigg)\Bigg(\prod_{a\in\mathsf{h}_{-}}\frac{\langle ap\rangle}{[aq]}\Bigg)\text{Pf}\,(\W{\Psi}_{\mathsf{h}})\,,
\end{equation}
where $\W{\Psi}_{\mathsf{h}}$ is composed of the following blocks in the same way as \eqref{eq:Psi}:
\begin{itemize}
\item $A$-part:
\begin{align}
	&\W{A}_{ab}=\frac{[ab]}{\langle ap\rangle\langle bp\rangle}\,.
\Label{eq:tA}
\end{align}	
\item $B$-part:
\begin{align}
	& a\in\mathsf{h}_{-}:& &\W{B}_{ab}=\left\{\begin{array}{>{\displaystyle}c @{\hspace{1em}} >{\displaystyle}l}
	\frac{\langle ap\rangle[bq]}{\langle ab\rangle} & b\in\mathsf{h}_{+}\\
	0 & b\in\mathsf{h}_{-}
	\end{array}\right.\,,& & a\in\mathsf{h}_{+}:& &\W{B}_{ab}=\left\{\begin{array}{>{\displaystyle}c @{\hspace{1em}} >{\displaystyle}l}
	\frac{[aq]\langle bp\rangle}{\langle ab\rangle} & b\in\mathsf{h}_{-}\\
	0 & b\in\mathsf{h}_{+}
	\end{array}\right.\,.
\Label{eq:tB}
\end{align}
\item $C$-part:
\begin{align}	
	& a\in\mathsf{h}_{-}:& &\W{C}_{ab}=\frac{[bq]}{\langle bp\rangle}\,, & & a\in\mathsf{h}_{+}:& &\W{C}_{ab}=\left\{\begin{array}{>{\displaystyle}c @{\hspace{1em}\vspace{0.2em}} >{\displaystyle}l}
	\frac{[ab]}{\langle ab\rangle} & b\neq a \\
	-\sum_{l\neq a}^{n}\frac{[al]\langle lq\rangle\langle lp\rangle}{\langle al\rangle\langle aq\rangle\langle ap\rangle} &b=a
	\end{array}\right.\,.
\Label{eq:tC}
\end{align}
\end{itemize}
We immediately observe that in $\W{C}$ the rows that belong to $\mathsf{h}_{-}$ are identical. Using this feature, we can perform the following elementary transformations, after choosing one reference particle $i\in\mathsf{h}_{-}$:
\begin{enumerate}
\item For all $j\in\mathsf{h}_{-}$ and $j\neq i$, subtract the $(s+j)$-th row and column of $\W{\Psi}_{\mathsf{h}}$ by the $(s+i)$-th row and column. This operation makes the $j$-th row of $\W{C}$ and the $j$-th column of $-\W{C}^{T}$ zero.
\item Subtract the first $s$ rows and columns (except for the $i$-th) of $\W{\Psi}_{\mathsf{h}}$ by a multiple of the $i$-th row (denoted by the subscript $i\times$) and column (denoted by the subscript $\times i$):
\begin{align*}
	&(\W{\Psi}_{\mathsf{h}})_{\times b}\rightarrow(\W{\Psi}_{\mathsf{h}})_{\times b}-(\W{\Psi}_{\mathsf{h}})_{\times i}\frac{\langle ip\rangle[bq]}{[iq]\langle bp\rangle}\,,\\
	&(\W{\Psi}_{\mathsf{h}})_{a\times}\rightarrow(\W{\Psi}_{\mathsf{h}})_{a\times}-(\W{\Psi}_{\mathsf{h}})_{i\times}\frac{\langle ip\rangle[aq]}{[iq]\langle ap\rangle}\,.
\end{align*}
This operation makes the $i$-th row of $\W{C}$ and the $i$-th column of $-\W{C}^{T}$ zero, except for $\W{C}_{ii}$. Less obviously, it also makes $\W{A}$ zero, except for the $i$-th row and column:
\begin{equation*}
\W{A}_{ab}=\frac{[ab]}{\langle ap\rangle\langle bp\rangle}\rightarrow\frac{[ab]}{\langle ap\rangle\langle bp\rangle}-\frac{[ai][bq]}{[iq]\langle ap\rangle\langle bp\rangle}-\frac{[aq][ib]}{[iq]\langle ap\rangle\langle bp\rangle}=0\,,
\end{equation*}
where the Schouten identity has been used in the numerator.
\item Subtract the first $s$ rows and columns of $\W{\Psi}_{\mathsf{h}}$ by a multiple of the $(s+i)$-th row and column:
\begin{align*}
	&(\W{\Psi}_{\mathsf{h}})_{\times b}\rightarrow(\W{\Psi}_{\mathsf{h}})_{\times b}-(\W{\Psi}_{\mathsf{h}})_{\times,s+i}\frac{[bi]}{[iq]\langle bp\rangle}\,,\\
	&(\W{\Psi}_{\mathsf{h}})_{a\times}\rightarrow(\W{\Psi}_{\mathsf{h}})_{a\times}-(\W{\Psi}_{\mathsf{h}})_{s+i,\times}\frac{[ai]}{[iq]\langle ap\rangle}\,.	
\end{align*}
This operation makes the entire $\W{A}$ zero.
\item Finally, for $j\in\mathsf{h}_{+}$, subtract the $(s+j)$-th row and column by a multiple of the $(s+i)$-th:
\begin{align*}
	&(\W{\Psi}_{\mathsf{h}})_{\times,s+j}\rightarrow(\W{\Psi}_{\mathsf{h}})_{\times,s+j}-(\W{\Psi}_{\mathsf{h}})_{\times,s+i}\frac{\langle ip\rangle[ji]}{[iq]\langle ji\rangle}\,,\\
	&(\W{\Psi}_{\mathsf{h}})_{s+j,\times}\rightarrow(\W{\Psi}_{\mathsf{h}})_{s+j,\times}-(\W{\Psi}_{\mathsf{h}})_{s+i,\times}\frac{\langle ip\rangle[ji]}{[iq]\langle ji\rangle}\,.	
\end{align*}
Now the $(s+i)$-th row and column of $\W{\Psi}_{\mathsf{h}}$ are zero except for $\W{C}_{ii}=[iq]/\langle ip\rangle$.
\end{enumerate}
We can then delete this row and column after pulling $\W{C}_{ii}$ out of the pfaffian, and we call the left-over $2s-2$ dimensional submatrix $\W{\Psi}'_{\mathsf{h}}$:
\begin{equation}
	\text{Pf}\,(\W{\Psi}_{\mathsf{h}})=(-1)^{s}\frac{[iq]}{\langle ip\rangle}\text{Pf}\,(\W{\Psi}'_{\mathsf{h}})\,,
\Label{eq:PftPsi}
\end{equation}
where in $\W{\Psi}'_{\mathsf{h}}$ the upper left diagonal block is identically zero, with dimension $s+s^{-}-2$. The lower left off-diagonal blocks has dimension $s^{+}\times (s+s^{-}-2)$, in which the columns are more than the rows when $s^{-}\geqslant 2$ and $s\geqslant 3$. Therefore, we can always find an elementary transformation to make at least one column zero in this block, such that $\text{Pf}\,(\W{\Psi}'_{\mathsf{h}})=\det(\W{\Psi}'_{\mathsf{h}})=0$. Finally, for the simplest case $s=s^{-}=2$, the last two lines and columns of $\W{\Psi}_{\mathsf{h}}$ are proportional since $\W{B}=0$, which makes $\text{Pf}\,(\W{\Psi}_{\mathsf{h}})$ vanish.

Therefore, we have proved that $\text{Pf}\,[\Psi_{\mathsf{h}}(\sigma)]=0$ for all possible $s$ with at least two negative helicity gravitons. As a result, the $(h^{-}h^{-})$ amplitude of EYM vanishes. Using the CHY formalism, we are now able to give an direct proof to the statement in~\cite{Bern:1999bx}: \emph{the gluon all-plus single-trace MHV amplitudes of EYM are identically zero.}

Before we proceed, we note that this technique can also be used to prove that $\text{Pf}\,'[\Psi(\sigma)]=0$ for all non-MHV amplitudes. The proof will be given in Appendix~\ref{sec:special}.

\subsection{Summary of Results}
Here we first give a summary on what we have done in this section. The CHY formalism states that the color-ordered EYM amplitude with $s$ gravitons and $r$ gluons can be obtained from:
\begin{equation}
	M(h_{1}\cdots h_{s},g_{{1}}\cdots g_{{r}})=\sum_{\{\omega\}\in\text{sol.}}\frac{1}{\omega_{g_{1}g_{2}}\omega_{g_{2}g_{3}}\ldots\omega_{g_{r}g_{1}}}\frac{\text{Pf}\,(\Psi_{\mathsf{h}})\text{Pf}\,'(\Psi)}{{\det}'(\Phi)}\,.
\end{equation}
At MHV, $\text{Pf}\,'(\Psi)$ is only supported by the special solution $\sigma$. By calculating $\text{Pf}\,[\Psi_{\mathsf{h}}(\sigma)]$, we find that:
\begin{itemize}
\item the $(g^{-}g^{-})$ and $(h^{-}g^{-})$ MHV amplitudes can be written as:
\begin{equation}
	M(h_{1}^{+}\cdots i^{-}\cdots g_{j}^{-}\cdots g_{r}^{+})\propto\frac{\langle ig_j\rangle^{4}}{\langle g_{1}g_{2}\rangle\langle g_{2}g_{3}\rangle\ldots\langle g_{r}g_{1}\rangle}\det(\phi_{\mathsf{h}_{+}})\,.
\Label{eq:MHVGeneral}
\end{equation}
See \eqref{eq:Mgghg} for details.
\item the $(h^{-}h^{-})$ MHV amplitude vanishes, due to $\text{Pf}\,[\Psi_{\mathsf{h}}(\sigma)]=0$, which analytically justifies the very well motivated conjecture given in~\cite{Bern:1999bx}.
\end{itemize}
In Sec.~\ref{sec:ConnectionSBDW}, we we will prove the equivalence between \eqref{eq:MHVGeneral} and the SBDW formula.

\section{The Connection to the SBDW Formula}\label{sec:ConnectionSBDW}

In \cite{Selivanov:1997aq,Selivanov:1997ts}, Selivanov proposed an exponential formula for tree-level single trace  $(g^-g^-)$ MHV amplitudes. Bern, De Freitas and Wong generalized KLT relation and then proposed that the $(g^-h^-)$ MHV amplitudes also have a similar exponential expression~\cite{Bern:1999bx}. In this paper, we refer this exponential formula as Selivanov-Bern-De Freitas-Wong (SBDW) formula, which states that:\footnote{Here we neglect an overall factor which depends on coupling constants.}
\begin{equation}
  M(h_{1}^{+}\cdots i^{-}\cdots g_{j}^{-}\cdots g_{r}^{+})\propto(-1)^{s^+}\frac{\langle ig_{j}\rangle^{4}}{\langle g_{1}g_{2}\rangle\langle g_{2}g_{3}\rangle\ldots\langle g_{r}g_{1}\rangle}S(\mathsf{h}_{+})\,,
\Label{eq:SBDW}
\end{equation}
where $S$ is given by:
\begin{align}
  S(\mathsf{h}_{+};\mathsf{p})&=\left.\left(\prod_{m\in\mathsf{h}_{+}}\frac{\partial}{\partial a_{m}}\right)\exp\left[\sum_{n_{1}\in\mathsf{h}_{+}}a_{n_{1}}\sum_{l\in\overline{\mathsf{h}_{+}}}\psi_{ln_{1}}\exp\left[\sum_{\substack{n_{2}\in\mathsf{h}_{+}\\n_{2}\neq n_{1}}}a_{n_{2}}\psi_{n_1n_2}\exp\left(\cdots\right)\right]\right]\right|_{a_{m}=0}\nonumber\\
&\equiv\left.\left(\prod_{m\in\mathsf{h}_{+}}\frac{\partial}{\partial a_{m}}\right)G(a_{\mathsf{h}_{+}};\mathsf{p})\right|_{a_m=0}\,,
\Label{eq:Sh+}
\end{align}
where $\overline{\mathsf{h}_{+}}$ is the complement of $\mathsf{h}_{+}$ in $\mathsf{p}=\{1,2,\ldots,n\}$. The matrix element $\psi_{ab}$ is related to $\phi_{ab}$ through:
\begin{equation}
  \psi_{ab}=\phi_{ab}\frac{\langle b\xi\rangle\langle b\eta\rangle}{\langle a\xi\rangle\langle a\eta\rangle}\,.
\Label{eq:refspinor}
\end{equation}
Our $S(\mathsf{h}_{+};\mathsf{p})$ is of a more general form comparing with the original one in~\cite{Bern:1999bx} since ours allows an arbitrary choice of the reference spinors. Using $\psi_{ab}$, We can define another matrix:
\begin{align}
  &(W_{n})_{ab}=\left\{\begin{array}{>{\displaystyle}c @{\hspace{1em}} >{\displaystyle}l}
  -\psi_{ab} & a\neq b \\
  \sum_{{k=1,\,k\neq i}}^{n}\psi_{ak} & a=b
  \end{array}\right.\,.
\end{align}
One then immediately sees that all the diagonal minors of $W_n$ are equal to those of $\phi$, independent of the choice of $|\xi\rangle$ and $|\eta\rangle$. In particular, we have:
\begin{equation*}
  \det(\phi_{\mathsf{h}_{+}})=(-1)^{s^{+}}\det[(W_n)_{\mathsf{h}_{+}}]\,.
\end{equation*}
Then to prove the equivalence of \eqref{eq:SBDW} and \eqref{eq:MHVGeneral}, we only need to show that:
\begin{equation}
  S(\mathsf{h}_{+};\mathsf{p})=\det[(W_n)_{\mathsf{h}_{+}}]\,.
\Label{eq:lemma}
\end{equation}
We will show that the both sides have the same graph theory interpretation and thus equal. 

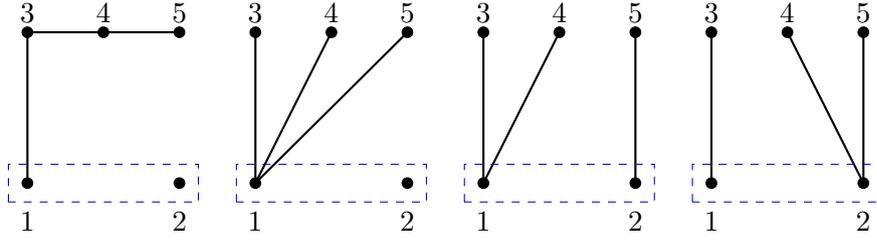
\begin{figure}[t]
\centering
  \begin{tikzpicture}
    \begin{scope}[xshift=0]
      \coordinate (A) at (0,0);
      \draw [fill=black] (A) circle (2pt) node [below=0.25cm]{$1$};
      \coordinate (B) at (2,0);
      \draw [fill=black] (B) circle (2pt) node [below=0.25cm]{$2$};
      \coordinate (C) at (0,2);
      \draw [fill=black] (C) circle (2pt) node [anchor=south]{$3$};
      \coordinate (D) at (1,2);
      \draw [fill=black] (D) circle (2pt) node [anchor=south]{$4$};
      \coordinate (E) at (2,2);
      \draw [fill=black] (E) circle (2pt) node [anchor=south]{$5$};
      \draw [thick] (A) -- (C) -- (D) -- (E);
      \draw [dashed,blue] (-0.25,-0.25) rectangle (2.25,0.25);
    \end{scope}
    \begin{scope}[xshift=3cm]
      \coordinate (A) at (0,0);
      \draw [fill=black] (A) circle (2pt) node [below=0.25cm]{$1$};
      \coordinate (B) at (2,0);
      \draw [fill=black] (B) circle (2pt) node [below=0.25cm]{$2$};
      \coordinate (C) at (0,2);
      \draw [fill=black] (C) circle (2pt) node [anchor=south]{$3$};
      \coordinate (D) at (1,2);
      \draw [fill=black] (D) circle (2pt) node [anchor=south]{$4$};
      \coordinate (E) at (2,2);
      \draw [fill=black] (E) circle (2pt) node [anchor=south]{$5$};
      \draw [thick] (D) -- (A) -- (C) (A) -- (E);
      \draw [dashed,blue] (-0.25,-0.25) rectangle (2.25,0.25);
    \end{scope}
    \begin{scope}[xshift=6cm]
      \coordinate (A) at (0,0);
      \draw [fill=black] (A) circle (2pt) node [below=0.25cm]{$1$};
      \coordinate (B) at (2,0);
      \draw [fill=black] (B) circle (2pt) node [below=0.25cm]{$2$};
      \coordinate (C) at (0,2);
      \draw [fill=black] (C) circle (2pt) node [anchor=south]{$3$};
      \coordinate (D) at (1,2);
      \draw [fill=black] (D) circle (2pt) node [anchor=south]{$4$};
      \coordinate (E) at (2,2);
      \draw [fill=black] (E) circle (2pt) node [anchor=south]{$5$};
      \draw [thick] (D) -- (A) -- (C) (B) -- (E);
      \draw [dashed,blue] (-0.25,-0.25) rectangle (2.25,0.25);
    \end{scope}
    \begin{scope}[xshift=9cm]
      \coordinate (A) at (0,0);
      \draw [fill=black] (A) circle (2pt) node [below=0.25cm]{$1$};
      \coordinate (B) at (2,0);
      \draw [fill=black] (B) circle (2pt) node [below=0.25cm]{$2$};
      \coordinate (C) at (0,2);
      \draw [fill=black] (C) circle (2pt) node [anchor=south]{$3$};
      \coordinate (D) at (1,2);
      \draw [fill=black] (D) circle (2pt) node [anchor=south]{$4$};
      \coordinate (E) at (2,2);
      \draw [fill=black] (E) circle (2pt) node [anchor=south]{$5$};
      \draw [thick] (C) -- (A) (D) -- (B) -- (E);
      \draw [dashed,blue] (-0.25,-0.25) rectangle (2.25,0.25);
    \end{scope}
  \end{tikzpicture}
\caption{\label{fig:forest}Some spanning forests of $K_5$ rooted in $I=\{1,2\}$.}
\end{figure}

First, the following identity holds for $S$:
\begin{align}
S(\overline{I}_{r};\mathsf{p})=\left.\left(\prod_{m\in\overline{I}_{r}}\frac{\partial}{\partial a_m}\right)G(a_{\overline{I}_r};\mathsf{p})\right|_{a_m=0}=\sum_{F\in\mathcal{F}_{I_r}(K_n)}\left(\prod_{v_av_b\in E(F)}\psi_{ab}\right)\,,
\Label{eq:lemma2}
\end{align}
where $K_{n}$ is a weighted complete graph\footnote{A complete graph $K_n$ is a simple graph (with no self-loop on the same vertex) in which each vertex is connected to the rest $n-1$ vertices. The graph $K_{n}$ is directed if $\psi_{ab}$ is not symmetric.} with the vertex set $\{v_{1},\ldots,v_{n}\}$ and weight $\psi_{ab}$ assigned to the edge $v_av_b$. The summation on the right hand side of (\ref{eq:lemma2}) is over $\mathcal{F}_{I_r}(K_n)$, the set of spanning forests of $K_n$ rooted on the vertices with labels $I_r=\{i_1,i_2,\ldots,i_r\}$. Some examples of spanning forests are presented in Fig.~\ref{fig:forest}. Such a forest $F$ has the same vertex set as $K_n$ while its edge set is denoted as $E(F)$. The forest $F$ has exactly $r$ trees, each of which contains one and only one vertex of $I_r$. Then $G(a_{\overline{I}_r};\mathsf{p})$, as defined in \eqref{eq:Sh+}, is a multivariate generating function of spanning forests in the sense that at fixed $n$ and $\psi_{ab}=1$, its Taylor expansion coefficient of $a_{i_{r+1}}a_{i_{r+2}}\cdots a_{i_n}$ equals the number of spanning forests of $K_{n}$ with $r$ trees rooted in $I_r$. In Sec.~\ref{sec:example}, we study a $7$-point example to help readers understand the idea of the general inductive proof, which will then be given in Sec.~\ref{sec:proof}. 

On the other hand, it has been demonstrated in~\cite{Feng:2012sy} that the evaluation of $\det[(W_n)_{\mathsf{h}_{+}}]$ has the same graph theory interpretation as $S$. The matrix $W_n$ actually represents the weighted complete graph $K_n$, while $\psi_{ab}$ is the weight assigned to the edge $v_{a}v_{b}$ and $(W_n)_{aa}$ is the total weight associated to the vertex $v_a$. We have a beautiful matrix-forest theorem\footnote{In~\cite{Feng:2012sy}, this theorem is called matrix-tree theorem II. We assume that all the edges are directed away from the roots.}, as given in~\cite{Feng:2012sy}:
\begin{equation}
  \det[(W_n)_{\overline{I}_r}]=\sum_{F\in\mathcal{F}_{I_r}(K_n)}\left(\prod_{v_av_b\in E(F)}\psi_{ab}\right)\,.
\Label{eq:statement}
\end{equation}
Thus if we choose $\mathsf{h}_{+}=\overline{I}_{r}$, the statement (\ref{eq:lemma}) is the direct consequence of our new theorem~(\ref{eq:lemma2}) and the matrix-forest theorem~(\ref{eq:statement}).
\subsection{A Seven-point Example of The Matrix-forest Theorem}
\label{sec:example}
In this section, we explicitly calculate the factor $S(\mathsf{h}_{+};\mathsf{p})$ of the amplitude:
\begin{equation*}
  M(h_1^+,h_2^+,h_3^+;g_1^+,g_2^+,g_3^-,g_4^-)\,,
\end{equation*}
and demonstrate the graphic correspondence of the expansion (\ref{eq:lemma2}). To simplify the presentation, we relabel the particles as:
\begin{align*}
  &h_{1}\equiv 1\,,& &h_{2}\equiv 2\,,& &h_{3}\equiv 3\,,& &g_{1}\equiv 4\,,& &g_{2}\equiv 5\,,& &g_{3}\equiv 6\,,& &g_{4}\equiv 7\,.
\end{align*}
The root set is $\{4,5,6,7\}$. However, by choosing the reference spinor in \eqref{eq:refspinor} to be:
\begin{align*}
  &|\xi\rangle=|g_3\rangle\equiv|6\rangle\,,& &|\eta\rangle=|g_4\rangle\equiv|7\rangle\,,
\end{align*}
we can substantially reduce the number of the graphs involved since now
\begin{equation*}
\psi_{6a}=\psi_{7a}=0\text{ for all }a\in\{1,2,3,4,5\}\,.
\end{equation*}
In other words, the vertex $6$ and $7$ are always disjoint to the others. We have thus reduced the problem to the spanning forests of $K_5$ (instead of $K_7$) with vertex set $\pmb{5}\equiv\{1,2,3,4,5\}$ and roots $\{4,5\}$. The SBDW generating function in this case is:
\begin{align}
G(\{1,2,3\};\pmb{5})=\exp\Big\{& a_1(\psi_{41}+\psi_{51})\exp\left[a_2\psi_{12}\exp(a_3\psi_{23})+a_3\psi_{13}\exp(a_2\psi_{32})\right]\nonumber\\
&+a_2(\psi_{42}+\psi_{52})\exp\left[a_3\psi_{23}\exp(a_1\psi_{31})+a_1\psi_{21}\exp(a_3\psi_{13})\right]\nonumber\\
&+a_3(\psi_{43}+\psi_{53})\exp\left[a_1\psi_{31}\exp(a_2\psi_{12})+a_2\psi_{32}\exp(a_1\psi_{21})\right]\Big\}\,.
\end{align}
Of course now one can expand
\begin{equation}
S(\{1,2,3\};\pmb{5})=\left.\frac{\partial}{\partial a_1}\frac{\partial}{\partial a_2}\frac{\partial}{\partial a_3}G(\{1,2,3\};\pmb{5})\right|_{a_1=a_2=a_3=0}
\Label{eq:S5}
\end{equation}
by brute force, enumerate the spanning forests, and demonstrate that \eqref{eq:lemma2} indeed holds in this case. Instead, we use a strategy that helps to better understand the inductive proof in the next subsection: we divide the graphs into lower order ones, check the theorem and then add them up.

First, we pick up the root $5$ and expand the right hand side of \eqref{eq:lemma2} with respect to those $\psi$'s whose subscripts contain $5$:
\begin{equation}
	\sum_{F\in\mathcal{F}_{\{4,5\}}(K_5)}\left(\prod_{v_av_b\in E(F)}\psi_{ab}\right)=A+\sum_{l\in\{1,2,3\}}\psi_{5l}B_{l}+\sum_{\{l,k\}\in\{1,2,3\}}\psi_{5l}\psi_{5k}C_{lk}+\psi_{51}\psi_{52}\psi_{53}D\,.
\Label{eq:example5p}
\end{equation}
In this way, we have put the spanning forests into four categories $\pmb A$, $\pmb B$, $\pmb C$ and $\pmb D$. The last term, $\psi_{51}\psi_{52}\psi_{53}D$, corresponds to the graph in which $1$, $2$ and $3$ are all connected to $5$:
\begin{equation}
\psi_{51}\psi_{52}\psi_{53}D=\;\adjustbox{raise=-0.75cm}{\begin{tikzpicture}
	\draw [very thick,fill=black] (-1,0) circle (2pt) node [anchor=north]{$4$};
	\draw [very thick,fill=black] (-1,1) circle (2pt) node [anchor=south]{$1$} -- (0,0) circle (2pt) node[anchor=north]{$5$} -- (0,1) circle (2pt) node[anchor=south]{$2$} (1,1) circle (2pt) node[anchor=south]{$3$} -- (0,0);
	\end{tikzpicture}
	}\;,
\end{equation}
which is the only possibility and $D=1$. In the second last term, $\psi_{5l}\psi_{5k}$ comes from the edges connecting $l$ and $k$ to $5$, while $C_{lk}$ is contributed by the spanning forests of the vertex set $\pmb{4}\equiv\{1,2,3,4\}$ with roots $\{l,k,4\}$. For example, for $l=2$ and $k=3$, the term $\psi_{52}\psi_{53}C_{23}$ corresponds to:
\begin{align}
\psi_{52}\psi_{53}C_{23}=\adjustbox{raise=-0.75cm}{\begin{tikzpicture}
	\draw [very thick,fill=black] (0,1) -- (0,0) circle (2pt) node [anchor=north]{$5$} -- (1,1);
	\filldraw [thick,blue] (-1,0) circle (2pt) node[anchor=north]{$4$} -- (-1,1) circle (2pt) node [anchor=south]{$1$} (0,1) circle (2pt) node [anchor=south]{$2$} (1,1) circle (2pt) node[anchor=south]{$3$};
	\end{tikzpicture}	
}
+
\adjustbox{raise=-0.75cm}{\begin{tikzpicture}
	\draw [very thick,fill=black] (0,1) -- (0,0) circle (2pt) node [anchor=north]{$5$} -- (1,1);
	\filldraw [thick,blue] (-1,0) circle (2pt) node[anchor=north]{$4$} (-1,1) circle (2pt) node [anchor=south]{$1$} -- (0,1) circle (2pt) node [anchor=south]{$2$} (1,1) circle (2pt) node[anchor=south]{$3$};
	\end{tikzpicture}	
}
+
\adjustbox{raise=-0.75cm}{\begin{tikzpicture}
	\draw [very thick,fill=black] (0,1) -- (0,0) circle (2pt) node [anchor=north]{$5$} -- (1,1);
	\filldraw [thick,blue] (-1,0) circle (2pt) node[anchor=north]{$4$} (-1,1) circle (2pt) node [anchor=south]{$1$}  (1,1) circle (2pt) node[anchor=south]{$3$} (0,1) circle (2pt) node [anchor=south]{$2$};
	\draw [thick,blue] (-1,1) .. controls (0,1.7) and (0,1.7) .. (1,1);
	\end{tikzpicture}	
}
=\psi_{52}\psi_{53}(\psi_{41}+\psi_{21}+\psi_{31})\,,
\end{align}
with the spanning forests of $\{1,2,3,4\}$ shown in blue. It is easy to see that
\begin{equation*}
G(\{1\};\pmb{4})=\exp\left[a_1\left(\psi_{21}+\psi_{31}+\psi_{41}\right)\right]
\end{equation*}
generates them all:
\begin{equation}
	C_{23}=\psi_{21}+\psi_{31}+\psi_{41}=S(\{1\};\pmb{4})=\left.\frac{d}{da_1}\exp\left[a_1\left(\psi_{21}+\psi_{31}+\psi_{41}\right)\right]\right|_{a_1=0}\,.
\end{equation}
Similar calculation applies to $C_{12}$ and $C_{13}$ such that there are in all $9$ graphs in the category $\pmb C$. With the experience gained from the previous calculations, one can immediately tells that in the second term of \eqref{eq:example5p}, $\psi_{5l}$ comes from the edge connecting $l$ and $5$ while $B_{l}$ is contributed by the spanning forests of $\pmb{4}$ with roots $\{l,4\}$. The result of $l=3$ is given below:
\begin{align}
\psi_{53}B_{3}&=\adjustbox{raise=-0.75cm}{\begin{tikzpicture}
	\draw [very thick,fill=black] (1,1) -- (0,0) circle (2pt) node [anchor=north]{$5$};
	\filldraw [thick,blue] (-1,0) circle (2pt) node[anchor=north]{$4$} (-1,1) circle (2pt) node [anchor=south]{$1$}  (1,1) circle (2pt) node[anchor=south]{$3$} (0,1) circle (2pt) node [anchor=south]{$2$};
	\draw [thick,blue] (0,1) -- (-1,0) -- (-1,1);
	\end{tikzpicture}	
}
+
\adjustbox{raise=-0.75cm}{\begin{tikzpicture}
	\draw [very thick,fill=black] (1,1) -- (0,0) circle (2pt) node [anchor=north]{$5$};
	\filldraw [thick,blue] (-1,0) circle (2pt) node[anchor=north]{$4$} (-1,1) circle (2pt) node [anchor=south]{$1$}  (1,1) circle (2pt) node[anchor=south]{$3$} (0,1) circle (2pt) node [anchor=south]{$2$};
	\draw [thick,blue] (-1,0) -- (-1,1) -- (0,1);
	\end{tikzpicture}	
}
+
\adjustbox{raise=-0.75cm}{\begin{tikzpicture}
	\draw [very thick,fill=black] (1,1) -- (0,0) circle (2pt) node [anchor=north]{$5$};
	\filldraw [thick,blue] (-1,0) circle (2pt) node[anchor=north]{$4$} (-1,1) circle (2pt) node [anchor=south]{$1$}  (1,1) circle (2pt) node[anchor=south]{$3$} (0,1) circle (2pt) node [anchor=south]{$2$};
	\draw [thick,blue] (-1,1) -- (0,1) -- (-1,0);
	\end{tikzpicture}	
}
+
\adjustbox{raise=-0.75cm}{\begin{tikzpicture}
	\draw [very thick,fill=black] (1,1) -- (0,0) circle (2pt) node [anchor=north]{$5$};
	\filldraw [thick,blue] (-1,0) circle (2pt) node[anchor=north]{$4$} (-1,1) circle (2pt) node [anchor=south]{$1$}  (1,1) circle (2pt) node[anchor=south]{$3$} (0,1) circle (2pt) node [anchor=south]{$2$};
	\draw [thick,blue] (-1,0) -- (-1,1) (0,1) -- (1,1);
	\end{tikzpicture}	
}\nonumber\\
&\quad+
\adjustbox{raise=-0.75cm}{\begin{tikzpicture}
	\draw [very thick,fill=black] (1,1) -- (0,0) circle (2pt) node [anchor=north]{$5$};
	\filldraw [thick,blue] (-1,0) circle (2pt) node[anchor=north]{$4$} (-1,1) circle (2pt) node [anchor=south]{$1$}  (1,1) circle (2pt) node[anchor=south]{$3$} (0,1) circle (2pt) node [anchor=south]{$2$};
	\draw [thick,blue] (-1,1) -- (0,1) -- (1,1);
	\end{tikzpicture}	
}
+
\adjustbox{raise=-0.75cm}{\begin{tikzpicture}
	\draw [very thick,fill=black] (1,1) -- (0,0) circle (2pt) node [anchor=north]{$5$};
	\filldraw [thick,blue] (-1,0) circle (2pt) node[anchor=north]{$4$} (-1,1) circle (2pt) node [anchor=south]{$1$}  (1,1) circle (2pt) node[anchor=south]{$3$} (0,1) circle (2pt) node [anchor=south]{$2$};
	\draw [thick,blue] (-1,1) .. controls (0,1.7) and (0,1.7) .. (1,1) -- (0,1);
	\end{tikzpicture}	
}
+
\adjustbox{raise=-0.75cm}{\begin{tikzpicture}
	\draw [very thick,fill=black] (1,1) -- (0,0) circle (2pt) node [anchor=north]{$5$};
	\filldraw [thick,blue] (-1,0) circle (2pt) node[anchor=north]{$4$} (-1,1) circle (2pt) node [anchor=south]{$1$}  (1,1) circle (2pt) node[anchor=south]{$3$} (0,1) circle (2pt) node [anchor=south]{$2$};
	\draw [thick,blue] (0,1) -- (-1,1) .. controls (0,1.7) and (0,1.7) .. (1,1);
	\end{tikzpicture}	
}
+
\adjustbox{raise=-0.75cm}{\begin{tikzpicture}
	\draw [very thick,fill=black] (1,1) -- (0,0) circle (2pt) node [anchor=north]{$5$};
	\filldraw [thick,blue] (-1,0) circle (2pt) node[anchor=north]{$4$} (-1,1) circle (2pt) node [anchor=south]{$1$}  (1,1) circle (2pt) node[anchor=south]{$3$} (0,1) circle (2pt) node [anchor=south]{$2$};
	\draw [thick,blue] (0,1) -- (-1,0) (-1,1) .. controls (0,1.7) and (0,1.7) .. (1,1);
	\end{tikzpicture}	
}\nonumber\\
&=\psi_{53}\left(\psi_{41}\psi_{42}+\psi_{41}\psi_{12}+\psi_{42}\psi_{21}+\psi_{41}\psi_{32}+\psi_{32}\psi_{21}+\psi_{32}\psi_{31}+\psi_{31}\psi_{12}+\psi_{31}\psi_{42}\right)\,.
\end{align}
We can verify that $B_3$ can also be generated by:
\begin{align}
G(\{1,2\};\pmb{4})=\exp\left[a_1(\psi_{31}+\psi_{41})\exp(a_2\psi_{12})+a_2(\psi_{32}+\psi_{42})\exp(a_1\psi_{21})\right]\,,
\end{align}
namely,
\begin{align}
S(\{1,2\};\pmb{4})&=\left.\frac{\partial}{\partial a_1}\frac{\partial}{\partial a_2}G(\{1,2\};\pmb{4})\right|_{a_1=a_2=0}\nonumber\\
&=\left.\frac{\partial}{\partial a_1}\left[a_1(\psi_{31}+\psi_{41})\psi_{12}+(\psi_{32}+\psi_{42})e^{a_1\psi_{21}}\right]e^{a_1(\psi_{31}+\psi_{41})}\right|_{a_1=0}\nonumber\\
&=(\psi_{31}+\psi_{41})\psi_{12}+(\psi_{32}+\psi_{42})\psi_{21}+(\psi_{31}+\psi_{41})(\psi_{32}+\psi_{42})\nonumber\\
&=B_{3}\,.
\end{align}
Including $B_1$ and $B_2$, which can be calculated similarly, we then have all the $24$ graphs in the category $\pmb B$. Finally, the category $\pmb A$ consists of those graphs with $5$ standing alone and $\{1,2,3,4\}$ connected, which are just the spanning tree of $\pmb{4}$. Thus there are in all $16$ graphs\footnote{The Cayley theorem states that the number of spanning trees of $K_n$ is $n^{n-2}$.} in the category $\pmb A$. It is straightforward, although tedious, to verify that the generating function:
\begin{equation*}
	G(\{1,2,3\};\pmb{4})=\left.G(\{1,2,3\};\pmb{5})\right|_{\psi_{51}=\psi_{52}=\psi_{53}=0}
\end{equation*}
indeed gives the correct value of $A$:
\begin{equation}
A=S(\{1,2,3\};\pmb{4})=\left.\frac{\partial}{\partial a_1}\frac{\partial}{\partial a_2}\frac{\partial}{\partial a_3}G(\{1,2,3\};\pmb{4})\right|_{a_1=a_2=a_3=0}\,.
\end{equation}
At the moment, we have verified that Theorem (\ref{eq:lemma2}) holds for all $4$-vertex spanning forests. Finally, one can verify by brute force that the expansion (\ref{eq:example5p}) indeed equals \eqref{eq:S5} such that Theorem (\ref{eq:lemma2}) also holds for our $5$-vertex forests. Astute readers may immediate see that if the final step holds, we have just completed an inductive-style proof from $4$-vertex graphs to $5$-vertex ones. The general inductive proof from level $n-1$ to level $n$ uses exactly the same construction.

To close this subsection, we note that in this example we have encountered an evaluation of
\begin{equation*}
|\pmb{A}|+|\pmb{B}|+|\pmb{C}|+|\pmb{D}|=16+24+9+1=50
\end{equation*}
forests, which keep proliferating for larger $n$. Thus without Theorem (\ref{eq:lemma2}), the SBDW-style calculation would be very involved and it would be very difficult to tell that such a massive summation over $50$ terms simply equals a much simpler $3\times 3$ determinant according to (\ref{eq:statement}). This example also demonstrates that in this case the CHY-style direct evaluation of EYM is much more powerful than the SBDW formula.
\subsection{General Proof}\label{sec:proof}
In this subsection, we present the general inductive proof of \eqref{eq:lemma2}. The method is parallel to the one used in~\cite{Feng:2012sy} to prove \eqref{eq:statement}. To start the induction, we first show that the $n=2$ case holds. The only nontrivial scenario at $n=2$ is $\overline{I}=\{1\}$ and $I=\{2\}$ (or vice versa), such that:
\begin{equation}
  \adjustbox{raise=-0.75cm}{\begin{tikzpicture}
      \draw [thick] (0,0) -- (0,1.5) node [pos=0.5,anchor=east]{$\psi_{12}$};
      \fill [black] (0,0) circle (2pt) (0,1.5) circle (2pt);
      \node at (0,0) [anchor=west]{$v_{1}$};
      \node at (0,1.5) [anchor=west]{$v_{2}$};
  \end{tikzpicture}}=\det[(W_2)_{\{1\}}]=\psi_{12}=S(\{1\};\{1,2\})=\left.\frac{d}{da_{1}}\exp\left(a_{1}\psi_{12}\right)\right|_{a_1=0}\,.
\end{equation}
Next, we assume that \eqref{eq:lemma2} holds for $n-1$. At $n$ vertices, we can construct a forest with roots $I_r$ by first constructing a forest of $n-1$ vertices with roots $I_{r-1+t}=\{i_{1},\ldots,i_{r-1},p_{1},\ldots,p_{t}\}$, and then connecting the vertex $i_r$ to $P_t=\{p_{1},\ldots,p_{t}\}\subset\overline{I}_r$:
\begin{equation}
  \sum_{F}\adjustbox{raise=-1.25cm}{\begin{tikzpicture}
      \foreach \x in {1,2,3} {
        \draw [fill=black] (\x-1,0) circle (2pt) node [anchor=north]{$i_{\x}$};
      }
      \node at (3,0) {$\cdots$};
      \draw [fill=black] (4,0) circle (2pt) node [anchor=north]{$i_r$};
      \draw [dashed,blue] (4.5,1) rectangle (-0.5,-1) node [anchor=south west]{$I_r$};
      \draw [thick] (-0.5,1.5) -- (0,0) -- (0.5,1.5) (1,0) -- (1,1.5) (1.5,1.5) -- (2,0) -- (2.5,1.5) (3.5,1.5) -- (4,0) -- (4,1.5) (4,0) -- (4.5,1.5);
      \draw [dashed,black] (-0.75,-1.5) rectangle (4.75,2) node [anchor=north east]{$F$};
  \end{tikzpicture}}\,=\sum_{t=1}^{n-r}\sum_{P_t}\sum_{F}\,  \adjustbox{raise=-1.25cm}{\begin{tikzpicture}
      \foreach \x in {1,2} {
        \draw [fill=black] (\x-1,0) circle (2pt) node [anchor=north]{$i_{\x}$};
      }
      \node at (1.5,0) {$\cdots$};
      \draw [fill=black] (2,0) circle (2pt) node [anchor=north]{$i_{r-1}$};
      \draw [fill=black] (4,0) circle (2pt) node [anchor=north]{$i_r$};
      \foreach \x in {3,3.5,4,5} {
        \draw [fill=black] (\x,0.75) circle (2pt) -- (\x,1.5);
        \draw [dashed,thick] (\x,0.75) -- (4,0);
      }
      \node at (4.5,0.75) {$\cdots$};
      \node at (3,0.75) [anchor=north]{$p_1$};
      \node at (5,0.75) [anchor=north]{$p_t$};
      \draw [dashed,blue] (-0.5,-1) -- (3,-1) -- (3,0.3) -- (5.5,0.3) -- (5.5,1) -- (-0.5,1) -- cycle;
      \node at (-0.5,-1) [blue,anchor=south west]{$I_{r-1+t}$};
      \draw [thick] (-0.5,1.5) -- (0,0) -- (0.5,1.5) (1,0) -- (1,1.5) (2,0) -- (2,1.5);
      \draw [dashed,black] (-0.75,-1.5) rectangle (5.75,2) node [anchor=north east]{$F$};
  \end{tikzpicture}}\,.
\end{equation}
Consequently, we have:
\begin{align}
  \sum_{F\in\mathcal{F}_{I_r}(K_n)}\left(\prod_{v_av_b\in E(F)}\psi_{ab}\right)&=\sum_{t=0}^{n-r}\sum_{P_t}\left(\prod_{k=1}^{t}\psi_{i_rp_k}\right)\sum_{F\in\mathcal{F}_{I_{r-1+t}}(K_{n-1})}\left(\prod_{v_av_b\in E(F)}\psi_{ab}\right)\nonumber\\
&=S(\overline{I}'_{r-1};\mathsf{p}')+\sum_{t=1}^{n-r}\sum_{P_t}\left(\prod_{k=1}^{t}\psi_{i_rp_k}\right)S(\overline{I}'_{r-1+t};\mathsf{p}')\,,
\Label{eq:expansion}
\end{align}
where the second line is obtained according to our induction assumption. The set $\overline{I}'_{r-1+t}$ is the complement of $I_{r-1+t}$ in the set $\mathsf{p}'=\mathsf{p}\backslash\{i_r\}$. In the following, a bar with a prime always means the complement of a set in $\mathsf{p}'$. In particular, we have $\overline{I}_r=\overline{I}'_{r-1}$.

Our next job is to show that \eqref{eq:expansion} is nothing but the expansion of $S(\overline{I}_r;\mathsf{p})$ with respect to $\psi_{i_rp_k}$. First, we observe that $\psi_{i_rp_k}$ only appear in the out-most level of $G(a_{\overline{I}_{r}},\psi)$:
\begin{equation}
  \sum_{n_1\in\overline{I}_r}a_{n_1}\sum_{l\in I_r}\psi_{ln_1}=\sum_{n_1\in\overline{I}'_{r-1}}a_{n_1}\left(\psi_{i_rn_1}+\sum_{l\in I_{r-1}}\psi_{ln_1}\right)\,.
\Label{eq:level1}
\end{equation}
By setting all $\psi_{i_rp_k}$ to zero, we obtain the zero-th order of $S(\overline{I}_r)$:
\begin{equation}
  \left.\left(\prod_{m\in\overline{I}'_{r-1}}\frac{\partial}{\partial a_{m}}\right)\exp\left[\sum_{n_1\in\overline{I}'_{r-1}}a_{n_1}\sum_{l\in I_{r-1}}\psi_{ln_1}\exp\left(\cdots\right)\right]\right|_{a_m=0}=S(\overline{I}'_{r-1};\mathsf{p}')\,,
\end{equation}
which agrees with the first term in \eqref{eq:expansion}. At a generic order $t$ with $P_t=\{p_1,\ldots,p_t\}$, we have:
\begin{equation}
  \prod_{m\in\overline{I}_{r}}\frac{\partial}{\partial a_{m}}=\left(\prod_{m\in\overline{I}'_{r-1+t}}\frac{\partial}{\partial a_{m}}\right)\underline{\left(\prod_{k=1}^{t}\frac{\partial}{\partial a_{p_{k}}}\right)}\,.
\Label{eq:der}
\end{equation}
In the exponent part, we can leave only those $\psi_{i_rp_k}$ in $P_t$ be nonzero such that \eqref{eq:level1} further transforms into:
\begin{equation}
    \sum_{n_{1}\in\overline{I}_{r}}a_{n_{1}}\sum_{l\in I_{r}}\psi_{ln_{1}}=\sum_{n_{1}\in\overline{I}'_{r-1+t}}a_{n_{1}}\sum_{l\in I_{r-1}}\psi_{ln_{1}}+\sum_{k=1}^{t}a_{p_{k}}\sum_{l\in I_{r-1}}\psi_{lp_{k}}+\underline{\sum_{k=1}^{t}a_{p_{k}}\psi_{i_rp_{k}}}\,.
\Label{eq:level12}
\end{equation}
Then the $t$-th order term corresponds to acting the underlined derivatives of \eqref{eq:der} onto the underlined part of \eqref{eq:level12}, and then setting all $a_{p_k}=0$:
\begin{align}
  &\quad\text{the $t$-th order of }\left.\left(\prod_{k=1}^{t}\frac{\partial}{\partial a_{p_{k}}}\right)G(a_{\overline{I}_r};\mathsf{p})\right|_{a_{p_k}=0}\nonumber\\
&=\left(\prod_{k=1}^{t}\psi_{i_rp_k}\right)\exp\left[\sum_{n_1\in\overline{I}'_{r-1+t}}a_{n_1}\sum_{k=1}^{t}\psi_{p_{k}n_{1}}\exp\left(\sum_{\substack{n_2\in\overline{I}'_{r-1+t}\\n_2\neq n_1}}a_{n_2}\psi_{n_1n_2}\cdots\right)\right]\nonumber\\
&\quad\times\exp\left[\sum_{n_1\in\overline{I}'_{r-1+t}}a_{n_1}\sum_{l\in I_{r-1}}\psi_{ln_1}\exp\left(\sum_{\substack{n_2\in\overline{I}'_{r-1+t}\\n_2\neq n_1}}a_{n_2}\psi_{n_1n_2}\cdots\right)\right]\nonumber\\
&=G(a_{\overline{I}'_{r-1+t}};\mathsf{p}')\,.
\Label{eq:ordert}
\end{align}
All the other ways of distributing derivatives result in lower order terms in the expansion. We also note that (\ref{eq:ordert}) is the only $t$-th order term we can have with $P_t$ specified. These two observations indicate that given the set $P_t$, the highest order in the expansion of $\psi_{i_rp_k}\in P_t$ is $t$. Since $P_t$ cannot be larger than $\overline{I}_{r}$, the expansion of $S(\overline{I}_{r};\mathsf{p})$ must terminate at the order $n-r$. Therefore, we have:
\begin{align}
  \text{the $t$-th order of $S(\overline{I}_{r};\mathsf{p})$ with $P_t$ }&=\left.\left(\prod_{k=1}^{t}\psi_{i_rp_k}\right)\left(\prod_{m\in\overline{I}'_{r-1+t}}\frac{\partial}{\partial a_{m}}\right)G(a_{\overline{I}'_{r-1+t}};\mathsf{p}')\right|_{a_m=0}\nonumber\\
&=\left(\prod_{k=1}^{t}\psi_{i_rp_k}\right)S(\overline{I}'_{r-1+t};\mathsf{p}')\,.
\end{align}
If we sum up all possible choices of $P_t$, we get:
\begin{equation}
  S(\overline{I}_{r};\mathsf{p})=S(\overline{I}'_{r-1};\mathsf{p}')+\sum_{t=1}^{n-r}\sum_{P_t}\left(\prod_{k=1}^{t}\psi_{i_rp_k}\right)S(\overline{I}'_{r-1+t};\mathsf{p}')\,,
\Label{eq:result}
\end{equation}
which is exactly \eqref{eq:expansion}.

The inductive proof of \eqref{eq:lemma2} is thus complete, and the desired equality:
\begin{equation}
  S(\mathsf{h}_{+};\mathsf{p})=\det[(W_n)_{\mathsf{h}_{+}}]=\det(\phi_{\mathsf{h}_{+}})
\end{equation}
follows immediatly by choosing $\overline{I}_{r}=\mathsf{h}_{+}$ in \eqref{eq:lemma2}. 

\section{Conclusion and Discussion}\label{sec:Conclusion}
In this paper, we proposed a new compact formula of the tree-level single-trace MHV amplitudes of Einstein-Yang-Mills theory, which results from a direct evaluation using the CHY formalism. The amplitudes with $(g^-g^-)$ and $(h^-g^-)$ configurations are expressed by multiplying a Parke-Taylor factor with a Hodges minor. We proved analytically that the amplitudes with  $(h^-h^-)$ configuration have to vanish. We also established a graph theoretical interpretation of the SBDW formula for MHV amplitudes of Einstein-Yang-Mills, and further proved that our new formula, \eqref{eq:MHVGeneral}, is equivalent to the SBDW formula.

There are some problems that still deserve further investigation:
\begin{itemize}
\item In the work \cite{Cachazo:2013iea}, CHY formula for amplitudes beyond single trace was discussed. Is there similar compact formula for more general amplitudes in EYM?
\item There are many discussions \cite{Chen:2009tr,Stieberger:2009hq,Chen:2010sr,Chen:2010ct,Chiodaroli:2014xia,Stieberger:2015qja,Stieberger:2015kia,Stieberger:2015vya,Chiodaroli:2015rdg,Stieberger:2016lng,Nandan:2016pya,delaCruz:2016gnm,Schlotterer:2016cxa} on the relation between EYM amplitudes and pure Yang-Mills amplitudes. How to relate this compact formula with those relations?
\end{itemize}

\section*{Acknowledgments}
YD would like to acknowledge National Natural Science Foundation of
China under Grant Nos. 11105118, 111547310, as well as the 351 program of Wuhan University.
\appendix

\section{The Special Solution and Non-MHV Amplitudes}
\label{sec:special}
In this section, we are going to prove that $\text{Pf}\,'[\Psi(\sigma)]=0$ for all non-MHV helicity configurations. This will settle a long standing conjecture that the special solution $\sigma$ only supports the MHV amplitudes, not any others~\cite{Cachazo:2013iaa,Monteiro:2013rya,Naculich:2014naa}. Now we can at least claim that the above statement is true for all the theories whose CHY integrand is proportional to $\text{Pf}\,'(\Psi)$. This class includes, but is not limited to, Yang-Mills, pure gravity and Einstein-Yang-Mills.

The proof flows almost parallel to the one in Sec.~\ref{sec:hh}. The quantity $\Psi$ is sensitive to helicity configurations, but not particle types, such that its structure resembles $\Psi_{\mathsf{h}}$ in Sec.~\ref{sec:hh} while the indices range within all external particles $\mathsf{p}$ instead of $\mathsf{h}$. The gauge choice in the polarizations are the same as that in \eqref{eq:gaugechoice}, but the index $a$ now ranges within $\mathsf{p}_{\pm}$ instead of $\mathsf{h}_{\pm}$. We plug in $\sigma$ given in \eqref{eq:solution} to $\Psi$, as given in \eqref{eq:Psi}, and then pull out common factors in rows and columns. We can then reach the result $\text{Pf}\,'[\Psi_{\mathsf{h}}(\sigma)]\propto\text{Pf}\,(\W{\Psi})$, where $\W{\Psi}$ is also given by (\ref{eq:tA}), (\ref{eq:tB}) and (\ref{eq:tC}). There are only two differences:
\begin{enumerate}
	\item there are two rows and columns deleted, as in \eqref{eq:rf}. Thus $\W{A}$ is now $(n-2)\times(n-2)$ dimensional and $\W{C}$ is $n\times(n-2)$ dimensional.
	\item all the indices now range within $\mathsf{p}_{\pm}$ instead of $\mathsf{h}_{\pm}$.
\end{enumerate}
Next, we perform the same set of elementary transformations as in Sec.~\ref{sec:hh} and we can also reach \eqref{eq:PftPsi}. In this case, the resultant matrix $\W{\Psi}'$ has a zero block with dimension $n-4+n^{-}$ in the upper left corner, and its lower left corner have dimension $n^{+}\times(n-4+n^{-})$. This block is square when $n^{-}=2$, which is exactly the MHV configuration, and the result has been given in \eqref{eq:Pfdet}. For non-MHV ($n^{-}\geqslant 3$), we have more columns than rows in $\W{\Psi}'$ such that we can always find an elementary transformation to make at least one column of the lower left corner zero, which leads to $\text{Pf}\,'[\Psi(\sigma)]\propto\text{Pf}\,(\W{\Psi}')=0$.

\bibliographystyle{JHEP}
\bibliography{Refs}

\end{document}